# High-temperature Ultraviolet Photodetectors: A Review

Ruth A. Miller,[1] Hongyun So,[2] Thomas A. Heuser,[3] and Debbie G. Senesky[1, a)]

[1]Department of Aeronautics and Astronautics

Stanford University, Stanford, CA 94305, USA

[2]Department of Mechanical Engineering

Hanyang University, Seoul 04763, Korea

[3]Department of Materials Science and Engineering

Stanford University, Stanford, CA 94305, USA

## Abstract

Wide bandgap semiconductors have become the most attractive materials in optoelectronics in the last decade. Their wide bandgap and intrinsic properties have advanced the development of reliable photodetectors to selectively detect short wavelengths (i.e., ultraviolet, UV) in high temperature regions (up to 300°C). The main driver for the development of high-temperature UV detection instrumentation is in-situ monitoring of hostile environments and processes found within industrial, automotive, aerospace, and energy production systems that emit UV signatures. In this review, a summary of the optical performance (in terms of photocurrent-to-dark current ratio, responsivity, quantum efficiency, and response time) and uncooled, high-temperature characterization of III-nitride, SiC, and other wide bandgap semiconductor UV photodetectors is presented.

a) Author to whom correspondence should be addressed. E-mail: dsenesky@stanford.edu

## I. INTRODUCTION

On the electromagnetic spectrum, the ultraviolet (UV) region spans wavelengths from 400 nm to 10 nm (corresponding to photon energies from 3 eV to 124 eV) and is typically divided into three spectral bands: UV-A (400–320 nm), UV-B (320–280 nm), and UV-C (280–10 nm). The Sun is the most significant natural UV source. The approximate solar radiation spectrum just outside Earth's atmosphere, calculated as black body radiation at 5800 K using Plank's law, is shown in Fig. 1. At the top of Earth's atmosphere, about 9% of the Sun's total radiation is in the UV regime [1]. Earth's ozone layer makes the planet habitable by humans by almost completely absorbing UV-C and significantly attenuating UV-B, which causes cataracts, burns, and skin cancer [2]. Artificial sources of UV radiation can be created by passing an electric current through a gas, typically mercury, causing energy to be released in the form of optical radiation. Further information on natural and artificial UV sources can be found in [1].

There are a wide range of ground and space-based applications that require UV detection including optical communication [3], flame detection [4], combustion monitoring [5], chemical analysis [6], astronomy [7], etc. Many of these applications require UV instrumentation capable of operating in high-temperature harsh environments. While most sensors and electronics today are silicon (Si)-based due to well-established manufacturing processes, easy circuit integration, and low cost, Si has shown limited usefulness as a high-temperature UV detecting material platform [8, 9]. Si and other narrow bandgap semiconductors are not able to directly measure UV light due to their bandgaps corresponding to near infrared or visible wavelengths. Since UV light is higher energy than the bandgap of Si and other narrow bandgap materials, part of the energy will be lost to heating, resulting in low quantum efficiency. To use these materials for UV photodetection, filtering devices that absorb UV light and reemit lower energy light need to be incorporated [10] or detectors with wide, shallow charge collection zones are required since the absorption depth of UV energy in Si is very small (less than 10 nm for wavelengths between 100–300 nm) [11, 12]. Additionally, Si-based photodetectors have been shown to be limited to temperatures less than about 125°C [8, 9]. Si has a relatively high intrinsic carrier concentration at room temperature ($10^{10}$ cm$^{-3}$) which increases exponentially with temperature to reach $10^{15}$ cm$^{-3}$ at 300°C [13]. Additionally, the lightest doping concentrations of Si devices range from $10^{14}$ to $10^{17}$ cm$^{-3}$ [14]. Therefore, when operating at high temperature, the intrinsic carrier concentration will overwhelm the doping concentration and the device will no longer function as intended.

Instead of Si, wide bandgap semiconductors such as III-nitrides, silicon carbide (SiC), and zinc oxide (ZnO) are being explored as material platforms for UV detection. The bandgaps of these materials allow for detection of UV light while remaining blind to visible light. Additionally, electronics based on wide bandgap materials have demonstrated high-temperature operation indicating their applicability to harsh environment sensing [14, 15]. A comparison of the material properties of Si and wide bandgap semiconductors used for UV detection is listed in Table I. Reviews of the latest progress in wide bandgap semiconductor UV photodetectors focusing on new device architectures and novel material combinations have been recently published [11, 16-27]. To complement and expand on these recent reviews, this review presents a summary of the high-temperature characterization and operation of UV photodetectors using wide bandgap semiconductors as well as the challenges that persist in furthering this technology for use in a variety of applications.

## II. PHOTODETECTOR PARAMETERS

There are several types of semiconductor device architectures that can be used to create photodetectors including photoconductor, metal-semiconductor-metal (MSM), Schottky, p-n, p-i-n, and avalanche. A schematic of each device architecture is shown in Fig. 2 and the theoretical framework that describes their operation can be found in [28-30]. Regardless of the photodetector architecture, the basic principal of operation of all semiconductor photodetectors is the same; photons with energy greater than or equal to the bandgap of the semiconductor are absorbed and generate electron-hole pairs. Under an applied electric field the electron-holes pairs separate and a photo-generated current can be measured. The main figures of merit that are used to quantify photodetector performance are photocurrent-to-dark current ratio (PDCR), responsivity (R), quantum efficiency ($\eta$), and response time.

PDCR is a measure of the photodetector sensitivity with respect to the dark (or leakage) current. PDCR is defined as

$$PDCR = \frac{I_{photo} - I_{dark}}{I_{dark}} \qquad (1)$$

where $I_{photo}$ is the measured photocurrent and $I_{dark}$ is the dark current. Responsivity is another measure of photodetector sensitivity and is defined as the photo-generated current ($I_{photo} - I_{dark}$) per unit of incident optical power, that is

$$R = \frac{I_{photo} - I_{dark}}{P_{opt}} \qquad (2)$$

where $P_{opt}$ is the applied optical power. Responsivity is directly proportional to the external quantum efficiency ($\eta$) which is a measure of the number of photo-generated electron-hole pairs per incident photon. The external quantum efficiency is defined as

$$\eta = R\frac{hc}{q\lambda} \qquad (3)$$

where $h$ is Plank's constant, $c$ is the speed of light, $q$ is elementary charge, and $\lambda$ is wavelength. Photodetector response time (which is related to bandwidth) is quantified in terms of photocurrent 10% to 90% rise time and photocurrent 90% to 10% decay time.

A comparison of the maximum operational temperatures reported in literature for Si, SiC, III-nitride, and other wide bandgap-based UV photodetectors is shown in Fig. 3. To date, SiC-based photodetectors have demonstrated the highest operational temperatures, up to 550°C [31]. The following sections detail the high temperature photodetector response in terms of PDCR, R, $\eta$, and response time for each of these material platforms. Additionally, the development and continuing challenges of using wide bandgap semiconductors for high temperature UV sensing are discussed.

### III. III-NITRIDE-BASED UV PHOTODETECTORS

The III-nitrides, consisting of GaN, InN, AlN, and their ternary compounds, present some unique benefits for UV photodetection when compared to other wide bandgap materials. III-nitride semiconductors have direct bandgaps and thus higher photon absorption than their indirect counterparts (such as Si or SiC). Also, the region of the electromagnetic spectrum, between 200 nm (AlN, bandgap of 6.2 eV) and 650 nm (InN, bandgap of 1.9 eV), that III-nitride detectors are sensitive to can be selected by changing the ternary compound mole fraction [11, 16]. Lastly, heterojunctions can be formed from III-nitride semiconductors and the highly conductive two-dimensional electron gas (2DEG) formed at the material interface can be leveraged as a photo-sensing element [32-35].

On the other hand, III-nitride materials have some limitations when used for UV sensing, the most significant of which is known as persistent photoconductivity (PPC). PPC is a phenomenon in which the photocurrent response remains after the illumination source has been removed as shown in Fig. 4, resulting in photocurrent decay times on the order of hours to days [32, 35-41]. PPC has been attributed to excitons [32], negatively charged surface states [38], metastable defects [37, 39, 40], gallium vacancies [36, 39], nitrogen antisites [40], and deep-level defects [39, 41] trapping photo-generated carriers. The falling transient is often fit by a stretched exponential function of the form

$$I_{PPC}(t) = I_0 \exp\left[-\left(\frac{t}{\tau}\right)^{\beta}\right]$$

(4)

where $I_0$ is the photocurrent before the illumination source is removed, $\tau$ is the decay time constant, and $\beta$ is the decay exponent ($0 < \beta < 1$) [36, 39-42]. Experimental results have shown that at elevated temperatures, thermal energy is able to release trapped photo-generated carriers and thus reduce PPC effects [32, 35, 37]. To account for this temperature dependence, the decay time constant has been modeled as

$$\tau = \tau_0 \exp\left(\frac{\Delta E}{kT}\right)$$

(5)

where $\Delta E$ is the carrier capture barrier, $k$ is Boltzmann's constant, and $T$ is the temperature [37, 40, 43]. The carrier capture barrier is thought to originate from the non-overlapping vibronic states of unfilled and filled defects [44]. To be captured, charge carriers require additional energy to get into the vibronic states of filled defects. Reported values for $\Delta E$ in III-nitride photodetectors range from 132 meV to 360 meV [35, 37, 40, 41]. This large energy barrier prevents the decay of photo-generated carriers at low temperatures. However, as trapped carriers gain thermal energy at elevated temperatures, the capture rate increases and thus the decay time is reduced.

In addition to increasing the photocurrent decay time, trapped charge carriers also affect photodetector quantum efficiency. Room temperature quantum efficiencies greater than 100% are often reported for III-nitride-based photodetectors and attributed to a photo-gain mechanism [38, 45-49]. The photo-gain has been attributed to photo-generated holes trapped at negatively-charged surface states at the metal-semiconductor interface causing the Schottky barrier to lower or bend with respect to the dark Schottky barrier as shown in Fig. 5 [38, 48, 49]. This change in the Schottky barrier results in an enhanced leakage current. Similar to PPC, thermal energy is able to release trapped photo-generated holes [46, 48]. Therefore, at elevated temperatures the change in the Schottky barrier from dark to illuminated conditions is less pronounced resulting in minimal photo-gain. In addition to a reduced photo-gain at high temperature, increased dark (or leakage) current, enhanced carrier recombination, increased lattice scattering, and shifting of the bandgap to longer wavelengths have been cited as reasons for lower quantum efficiencies, responsivities, and PDCRs at elevated temperatures [32, 33, 50, 51]. Increased responsivity at higher temperatures has also been reported and has been attributed to thermal ionization of trapped photo-generated carriers providing additional carriers [32, 45, 52].

Table II details the high temperature performance of III-nitride-based photodetectors reported in literature. The highest reported operational temperature of any III-nitride-based photodetector is 327°C by an AlGaN/GaN MSM device [32]. However, this photodetector exhibited decay times more than a factor of two greater than a similar MSM photodetector fabricated on a GaN substrate [32]. These long decay times were attributed to a high defect density as determined from surface roughness measurements. Thus, to reduce PPC effects, high quality III-nitride films are needed. Thicker GaN films have been shown to contain fewer defects and indeed, MSM photodetectors on a 4.0 μm thick GaN film showed improved performance compared to photodetectors fabricated on 1.5 μm thick GaN film [53].

Aside from growing higher quality, less defective III-nitride films, another proposed approach to mitigate PPC effects is to use in situ heating. Suspended AlGaN/GaN photodetectors utilizing the 2DEG as both a sensor and heater demonstrated a photocurrent decay time of 24 seconds which was more than three orders of magnitude less than the 39 hour decay time of their solid state counterpart [35]. The suspended photodetectors used joule heating of the 2DEG to accelerate the carrier capture rate once the UV source was removed which successfully mitigated PPC effects.

To increase quantum efficiency, PDCR, and responsivity at high temperatures, thermally stable contacts with low dark current are needed. To accomplish this, calcium fluoride (CaF$_2$) was tested as an insulation layer in InGaN-based metal-insulator-semiconductor (MIS) Schottky photodiodes [45, 52]. Compared to the commonly used SiO$_2$ insulation layer, CaF$_2$ has an extremely wide bandgap (12 eV), high thermal conductivity, and is robust to radiation [52]. The addition of the CaF$_2$ insulation layer decreased the reverse leakage current by three orders of magnitude compared to Schottky photodiodes without the insulation layer [45, 52]. The MIS device demonstrated operation up to 250°C with a PDCR values of 1317 at room temperature and 72 at 250°C [45, 52]. Whereas Schottky photodiodes without the insulation layer were only operational up to 200°C with lower PDCRs of 79 at room temperature and 2.2 at 200°C [45, 52].

Another approach to increase quantum efficiency, PDCR, and responsivity at high temperatures is to increase the UV light absorption. Three-dimensional AlGaN/GaN UV photodetectors using a "v-grooved" Si substrate demonstrated operation up to 200°C with improved PDCR over planar AlGaN/GaN UV photodetectors (~1.4 compared to ~0.9 at room temperature and ~0.3 compared to ~0.1 at 200°C ) [33]. The increased sensitivity was attributed to the increased absorption (via redirection of reflected light) of the incident UV light in the three-dimensional substrate. III-nitride photodetectors with ZnO nanorad arrays have also demonstrated enhanced sensitivity at high temperatures due to their anti-reflective properties [54]. Increased photodetector sensitivity due to ZnO nanorod arrays was demonstrated with a direct wirebond photodetector where a GaN chip was directly wirebonded to a carrier chip thus avoiding time consuming and costly microfabrication steps such photolithography, metal sputtering (or evaporation), and etching (or lift-off) [54]. These rapidly fabricated/packaged photodetectors had a PDCR of 1.11 at room temperature and demonstrated operation up to 250°C with a PDCR of 0.11. With the addition of an anti-reflective ZnO nanorod array coating to enhance light trapping, the direct wirebond photodetector sensitivity improved to 2.63 and 0.27 at room temperature and 250°C, respectively [54].

Lastly, few reports of AlN photodetectors have been published. However, AlN is a promising material candidate for high temperature deep-UV (< 200 nm) sensing applications. AlN has a bandgap of 6.2 eV which corresponds to a cut-off wavelength of 200 nm, making AlN photodetectors not only visible blind but also solar blind. MSM photodetectors fabricated on AlN thin films grown on a Si substrate demonstrated room temperature dark currents below 1 nA at applied bias voltages up to 200 V and room temperature PDCRs as high as 63 [55]. The AlN photodetectors demonstrated operation up to 300°C (PDCR of 3.5) and the room temperature response was fully recovered after thermal cycling up to 400°C [55].

## IV. SiC-BASED UV PHOTODETECTORS

SiC is another attractive wide bandgap semiconductor for UV detection at high temperatures. In particular, 4H-SiC has strong chemical bonds, high thermal conductivity (4−4.9 W·cm$^{-1}$·K$^{-1}$) [58], and high electron saturation velocity ($2.7\times10^7$ cm·s$^{-1}$) [59] enabling 4H-SiC based UV photodetectors to operate in high-temperature and high-radiation environments with fast response speed. Table III summarizes the type (structure) of device, electrode metal, operation temperature, PDCR, and responsivity of SiC-based UV photodetectors. The average operation temperature is slightly higher than the operation temperature of GaN-based UV photodetectors while the overall values of responsivity are relatively smaller than those of GaN-based photodetectors. The highest reported operational temperature of SiC-based photodetectors is 550°C by p-i-n structure [31]. From room temperature to 550°C, the photocurrent increased by 9 times at 365 nm and decreased by 2.6 times at 275 nm due to bandgap narrowing effect at high temperature as shown in Fig. 6 [31, 60]. This thermally-induced bandgap narrowing effect also induces an increase in the optical absorption coefficient [61, 62], thus increasing the quantum efficiency of photodetectors as temperature increases. Based on this effect, the highest quantum efficiency of 53.4% and 63.6% was reported at room temperature and 150°C, respectively, by 4H-SiC avalanche photodiodes [63]. For long-term reliability at high temperature, a 4H-SiC photodetector using Schottky metal contacts was exposed to 200°C in air for 100 hours [64].

After thermal storage, the responsivity and dark current level of the device remained unchanged, thus indicating the reliable operation of 4H-SiC UV photodetectors at high temperatures up to 200°C [64]. The temperature-independent responsivity of SiC photodetectors was also studied since the photoresponse usually depends on temperature. By controlling the reverse-bias voltage from 0 to 150 V, 4H-SiC p-n photodiodes were shown to have temperature-independent responsivity under 280−300 nm wavelength range as shown in Fig. 7 [65]. This phenomenon was explained by a combination of temperature-dependent optical absorption coefficient and surface recombination effects [65].

Compared to GaN-based photodetectors, severe PPC effect was not observed in SiC-based photodetectors which enables photodiodes to have a fast response time. The rise/fall time of an MSM photodetector using 7 μm p-type 4H-SiC epitaxial layer was reported as 594 μs/699 μs and 684 μs/786 μs at room temperature and 400°C, respectively [8]. Because the bandwidth of 4H-SiC is decreased due to the decrease in hole/electron saturation velocity as temperature increases, the rise/fall time of the photodetector is slightly increased at high temperature [8]. Although many SiC-based photodetectors are fabricated on 4H-SiC epitaxial layer, 6H-SiC [66], nanocrystalline SiC [67], and β-SiC (or 3C-SiC) [68] have also been used for fabrication of UV photodetectors. In particular, β-SiC on Si substrate was used to obtain high gain of optical sensor [68]. To extend the operation temperature and improve the high temperature performance, porous silicon substrate was additionally adopted as the semi-insulating substrate, suppressing the leakage current and thus resulting in a low dark current level [68]. The synthesis of ZnO nanorod arrays (i.e., antireflective coating) on SiC layer was reported as an alternative method to increase the operation temperature of the photodetectors [67].

## V. OTHER TYPES OF UV PHOTODETECTORS

In addition to GaN and SiC as discussed above, there are other less developed materials with bandgaps appropriate for ultraviolet photodetection that are also capable of operating at elevated temperatures including zinc oxide (ZnO), gallium oxide ($Ga_2O_3$), diamond, and boron nitride (BN) [24-27]. To date, a $Ga_2O_3$ MSM photodetector with transparent indium zinc oxide (IZO) electrodes has the highest reported operating temperature (427°C) of all photodetectors based on these other materials [71]. Table IV details the reported operating temperatures of photodetectors based on these materials with the benefits and drawbacks each material discussed further below.

ZnO, a II-VI semiconductor, is often compared to GaN. Both preferentially form in the wurtzite crystal structure, and they have direct band gaps with very similar energies (3.37 eV vs. 3.4 eV for GaN, corresponding to a cut-off wavelength of 368 nm) [72]. This, combined with a high decomposition temperature (~1975°C), means that ZnO is very well suited for use in high-temperature UV photodetectors. Although ZnO has been studied extensively as a material for UV photodetection, there has been little investigation into the behavior of ZnO-based devices at elevated temperatures, with the highest reported testing (of an MSM device with Al electrodes) carried out at 200°C [72]. Operation at this temperature showed reduced photocurrent and responsivity, as well as increased dark current compared with samples tested at room temperature. These changes were attributed to band gap shrinkage and increased lattice scattering [72].

$\beta$-$Ga_2O_3$, with a band gap of 4.9 eV (253 nm), is the most stable of the five polymorphs of $Ga_2O_3$. It has a melting point of ~1780°C ($\alpha$-$Ga_2O_3$ has a higher theoretical melting point of ~1900°C but transforms into $\beta$-$Ga_2O_3$ at temperatures over 800°C) [73]. Much of the interest in

β-Ga$_2$O$_3$ stems from its potential as a transparent conducting oxide (TCO) that allows for transmission of long-wavelength ultraviolet light, a result of its large band gap [74]. A variety of β-Ga$_2$O$_3$ UV photodetectors (mostly Schottky diodes, and MSM devices) have been tested at elevated temperatures, and are particularly notable for their short decay times, extremely low dark currents, and solar-blind photoresponses [71, 75-78]. β-Ga$_2$O$_3$ UV photodetectors have also demonstrated robustness through little variation in device behavior with changing atmospheric oxygen concentration and good resistance to permanent degradation at elevated temperatures, with one device (MSM with IZO electrodes) showing full recovery of room-temperature behavior after testing at 427°C, the current high-temperature record for β-Ga$_2$O$_3$ UV photodetectors [71].

Diamond has a band gap of ~ 5.5 eV (225 nm), is one of the hardest known materials, has a very high (pressure-dependent) theoretical sublimation temperature (over 3000°C), but readily transforms into graphite at temperatures above ~1200°C. (Diamond is thermodynamically unstable with respect to graphite at all temperatures, but the transformation is extremely slow at lower temperatures) [79]. Pure diamond is often doped (particularly with boron) to improve its semiconducting properties [80]. MSM photodetectors fabricated on 1 μm-thick polycrystalline CVD-deposited diamond films with Ag electrodes have been tested up to 300°C, displaying a responsivity of over 100 mA/W for the range from 225-350 nm [81].

Hexagonal boron nitride (h-BN) has a band gap of ~6 eV (206 nm) corresponding to the deep ultraviolet and is particularly notable, even among wide-band gap semiconductors, for its high decomposition temperature (2,973°C). Due to its extremely wide bandgap, h-BN is solar-blind and therefore does not need solar rejection filters to operate in the deep ultraviolet [82]. H-BN can be synthesized in a variety of forms including nanorods, nanotubes, and nanosheets (BNNS) with 2D nanosheets of special interest due to their physical and structural similarities (in addition to thermal stability) to graphene [83]. Photodetectors based on BNNS with Au electrodes tested up to 400°C displayed a factor of four increase in photocurrent and a factor of three increase in thermal noise compared to devices tested at room temperature [82, 83]. This combination of high photocurrent, solar-blind photoresponse, and high thermal stability mean that BNNS is particularly well suited to use in high-temperature ultraviolet photodetectors.

## VI. CONCLUSIONS

Wide bandgap semiconductors have advanced high temperature UV sensing beyond the capabilities of silicon due to their intrinsic thermal stability. Of the wide bandgap semiconductors used for high temperature UV detection, the III-nitrides and SiC are the most developed and have demonstrated operating temperatures upwards of 300°C. However, further development of all of these material platforms is needed to enable high UV sensitivity with long term reliable operation at elevated temperatures. For these materials to reach their full potential, material quality issues need to be addressed, high temperature capable electrode technology needs to be furthered, and high temperature packaging schemes need to be developed.

## AUTHOR INFORMATION

**Corresponding Authors**
*E-mail: dsenesky@stanford.edu

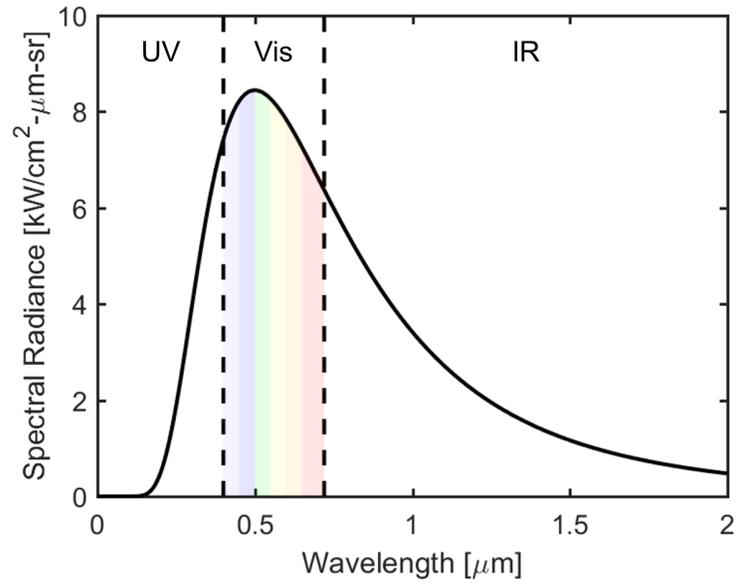

FIG 1. Black body radiation at 5800 K approximating the solar radiation spectrum outside Earth's atmosphere.

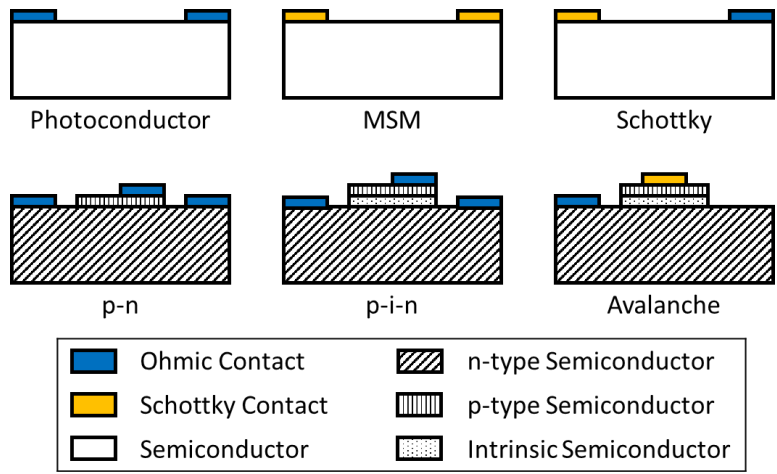

FIG 2. Different semiconductor photodetector device architectures.

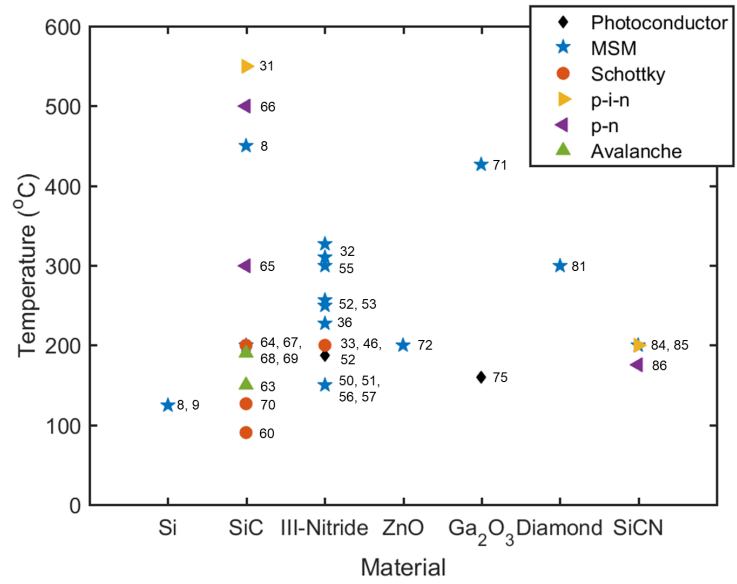

FIG 3. Maximum operational temperatures reported in literature for Si photodetectors compared to photodetectors based on wide bandgap materials.

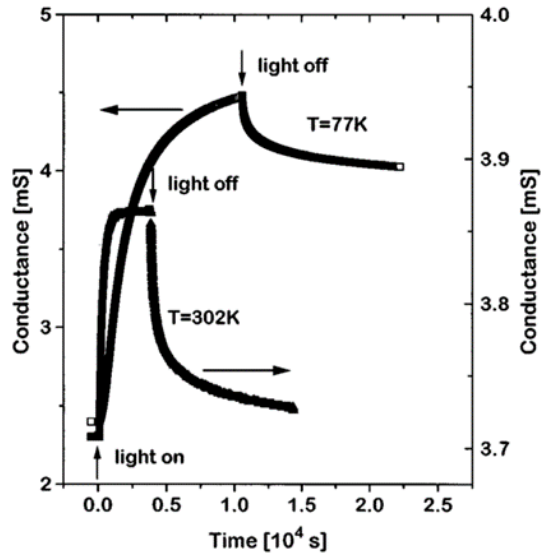

FIG 4. Buildup and decay transients of an n-type GaN substrate at 77 K and 302 K demonstrating decreased time constants at higher temperatures. Fig. 4 is adopted with permission from Ref. 37 (Copyright 1997, AIP Publishing).

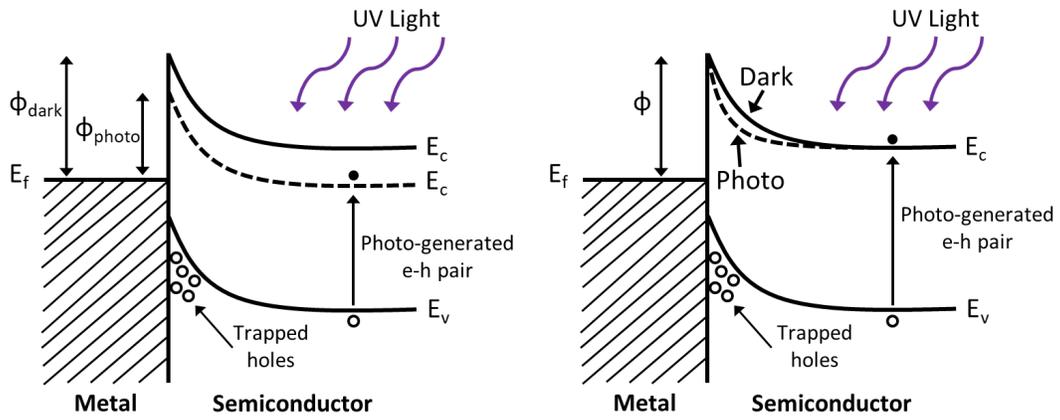

FIG 5. Trapped photo-generated holes lowering or bending the Schottky barrier with respect to the dark Schottky barrier.

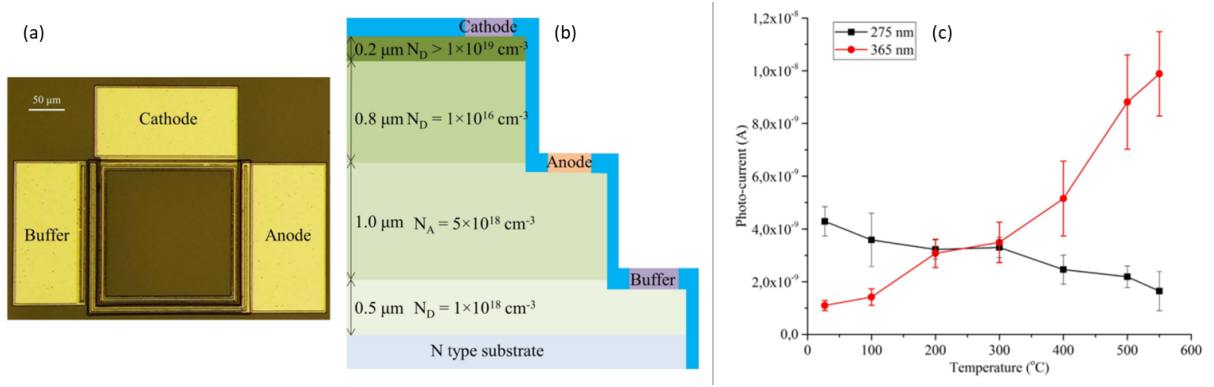

FIG 6. (a) Microscope image, (b) cross-section, and (c) photocurrent as a function of temperature for a SiC p-i-n photodetector. As temperature increases, the photocurrent generated by 275 nm light decreases while, the photocurrent generated by 365 nm light increases indicating bandgap narrowing at high temperature affects the photodetector response. Fig. 6 is adopted with permission from Ref. 31 (Copyright 2016, IEEE).

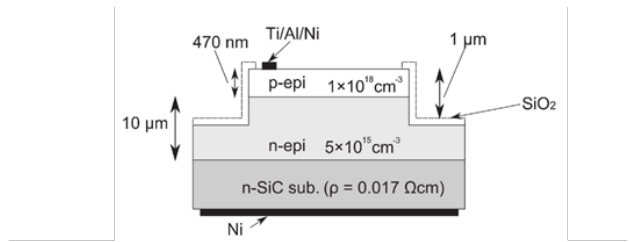

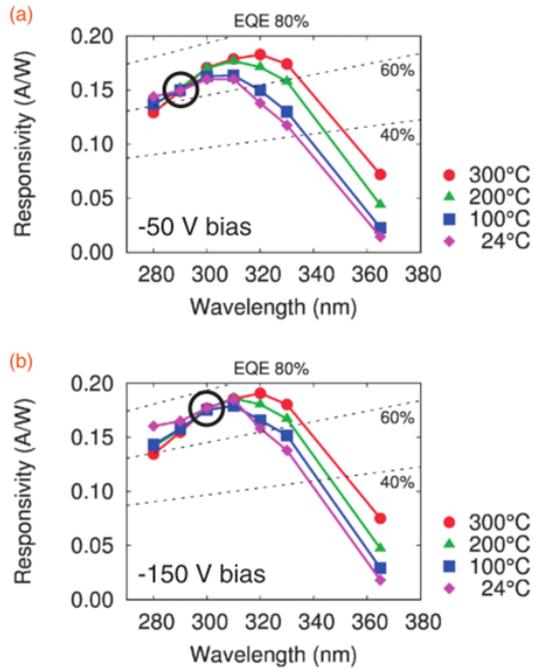

FIG 7. SiC p-n photodiodes demonstrating a temperature-independent photoresponse at a specific wavelength based on applied bias voltage. Fig. 7 is from Ref. 65 (Copyright 2012, IOP Publishing Ltd.).

TABLE I. Material properties of semiconductors used to manufacture UV photodetectors.

| | Si | AlN | Diamond | GaN | 4H-SiC | 6H-SiC | ZnO |
|---|---|---|---|---|---|---|---|
| **Bandgap (eV)** | 1.12 Indirect | 6.2 Direct | 5.5 Indirect | 3.4 Direct | 3.2 Indirect | 2.86 Indirect | 3.37 Direct |
| **Cut-off Wavelength (nm)** | 1107 | 200 | 225 | 365 | 387 | 434 | 368 |
| **Melting Point (°C)** | 1410 | 2200 | 3500 | 2500 | 2700 | 2700 | 1975 |
| **Thermal Conductivity (W/cm-K)** | 1.5 | 3.2 | 20 | 1.3 | 4.9 | 4.9 | 5.4 |
| **Electron Mobility (cm$^2$/V-s)** | 1400 | 135 | 2200 | 1000 | 400 | 400 | 205 |
| **Hole Mobility (cm$^2$/V-s)** | 600 | 14 | 1600 | 30 | 75 | 75 | 70 |
| **Dielectric Constant** | 11.8 | 8.1 | 5.5 | 8.9 | 9.7 | 9.7 | 9.1 |
| **Breakdown Field (10$^5$ V/cm)** | 3 | 20 | 100 | 26 | 24 | 24 | 35 |

TABLE II. High temperature III-nitride-based UV photodetectors.

| Material | Photodetector Architecture | Electrode Material | Temperature (°C) | PDCR | Responsivity (A/W) | Notes | Reference |
|---|---|---|---|---|---|---|---|
| GaN | MSM | Ti/Au | 310 | | ~8 (77°C) ~13 (310°C) | Thermal ionization of trapped photo-generated carriers increases responsivity. | 32 |
| GaN | MSM | Ni/Au | 150 | | ~0.1 (RT, 3V) ~0.04 (150°C, 3V) | Internal gain attributed to trapped photo-generated holes reducing the Schottky barrier height. | 56 |
| GaN | | Al Wirebond | 250 | 2.63 (RT, 1V) 0.27 (250°C, 1V) | | Direct wirebonding. ZnO nanorods for antireflective coating. | 54 |
| GaN | MSM | Ni/Au | 150 | | | Dark current ~ $10^{-12}$ A at RT and 150°C. Photocurrent ~$10^{-5}$ at RT and ~$10^{-6}$ at 150°C. | 50 |
| GaN | Photoconductor | Ti/Al or In | 187 | | | Very high gains (~$10^4$ at 187°C and $10^{-2}$ W/cm$^2$ optical power) attributed to a modulation mechanism of the conductive volume of the GaN layer. | 46 |
| GaN | MSM | Pt | 257 | | | Thicker GaN devices showed improved performance due to a reduction in defects and deep-level traps. | 53 |
| GaN | MSM | Ti/Au | 227 | | | Studied trapping mechanisms responsible for persistent photoconductivity. | 36 |

| AlGaN | MSM | Ni/Au | 150 | | 0.14 (RT, 10V) 0.11 (150°C, 10V) | Ultra-low dark current (fA range at 150°C) achieved using a high-temperature AlN buffer layer. | 51 |
|---|---|---|---|---|---|---|---|
| AlGaN | MSM | Ni/Au | 150 | | | QE: 11.3% (RT, 2V, front-illumination), 21.1% (RT, 2V, back-illumination) | 57 |
| InGaN | MIS | Ti/Al/Ni/Au Ni/WC | 250 | 1317 (RT, -1V) 72 (250°C, -1V) | 3.3 (RT, -3V) 5.6 (250°C, -3V) | UV/visible light discrimination ratio of $10^5$ at 250°C. Used $CaF_2$ insulation layer to reduce the dark current. | 45, 52 |
| InGaN | Schottky | Ti/Al/Ni/Au Ni/WC | 200 | 79 (RT, -1V) 2.2 (200°C, -1V) | | | 52 |
| AlGaN/GaN | MSM | Ti/Au | 327 | | ~1 (27°C) ~18 (327°C) | Thermal ionization of trapped photo-generated carriers increases responsivity. | 32 |
| AlGaN/GaN | Photoconductor | Ti/Al/Pt/Au | 200 | 1.4 (RT, 1V) 0.3 (200°C, 1V) | 0.00035 (RT and 200°C, 1V) | V-grooved photodetector increases absorption of incident light resulting in higher sensitivity compared to planar photodetector. | 33 |
| AlGaN/GaN | | Al Wirebond | 100 | | | Direct wirebonding enabling rapid fabrication and packaging. Signal to noise ratio: 29.2 (RT) and 14.4 (100°C) | 34 |

| AlGaN/GaN | | Ti/Al/Pt/Au | 270 (Membrane Heating) | 0.04 (RT, 30V) | | Photocurrent decay time: 39 hours (1 V = membrane temperature 25°C) and 24 seconds (30 V = membrane temperature 270°C) | 35 |
| AlN | MSM | Ti/Pt | 300 | 60 (RT, 5V) 3.5 (300°C, 5V) | 0.015 (RT, 5V) | Rise time ~110 ms Decay time ~ 80ms | 55 |

TABLE III. High temperature SiC-based UV photodetectors.

| Material | Photodetector Architecture | Electrode Material | Temperature (°C) | PDCR | Responsivity (A/W) | Notes | Reference |
|---|---|---|---|---|---|---|---|
| 4H-SiC | p-i-n | Ni Ni/Ti/Al | 550 | 45.4 (500°C) 7.3 (550°C) | 0.12 (RT) 0.046 (550°C) | From RT to 550°C, photocurrent increased by 9 times at 365 nm and decreased by 2.6 times at 275 nm due to bandgap narrowing. | 31 |
| 4H-SiC | MSM | Cr/Pd | 450 | 1.3 x 105 (RT) 0.62 (450°C) | 0.305 (RT, 20V) | Rise time: 594 μs (RT) and 684 μs (400°C) Fall time: 699 μs (RT) and 786 μs (400°C) | 8 |
| 4H-SiC | Schottky | Ti/AlSiCu Ti/Ni/Au | 90 | | 0.046 (RT) | QE: 19% (RT) | 60 |
| 4H-SiC | Avalanche | Ni/Ti/Al/Au | 150 | | 0.125 (RT) | Maximum QE: 53.4% (RT, 290 nm) and 63.3% (150°C, 295 nm) | 63 |
| 4H-SiC | Avalanche | Ni/Ti/Al/Au | 190 | | 0.093 (RT) | QE: 41% (RT) | 69 |
| 4H-SiC | p-n | Ti/Al/Ni | 300 | | Between 0.15 and 0.2 (RT and 300°C) | Achieved a temperature independent photoresponse at targeted wavelengths by controlling the reverse bias voltage. | 65 |
| 4H-SiC | Schottky | Ni/Ti/Al/Au Ni | 200 | | 0.115 (RT) | QE: 50% (RT) After thermal storage at 200°C in air for 100 hours, dark current increased slightly but remained less than 1 pA   at 20 V. | 64 |
| 4H-SiC | Schottky | Cr | 127 | | | QE: 26% (-198°C), 27.4% (102°C) | 70 |
| 6H-SiC | n+-p | Ni Ti/Al | 500 | | 0.0586 (RT) | QE: 24.9% (RT) Photocurrent increased with increasing temperature. | 66 |
| Nanocrystalline SiC | MSM | Au | 200 | RT: 4.9, 13.3 (with ZnO Nanorods) 200°C: 4.9, 7.6 (with ZnO nanorods) | | Used ZnO nanorod arrays as an antireflective coating. | 67 |
| β-SiC on Porous Si | MSM | Al | 200 | 30 (RT, 5V) 15 (200°C, 5V) | 0.28 (RT, 5V) 0.25 (200°C, 5V) | High resistivity and flexible porous silicon substrate results in low dark current and improved high temperature performance. | 68 |
| β-SiC on Si | MSM | Al | 200 | 8 (RT, 5V) 1(200°C, 5V) | 0.2 (RT, 5V) 0.17 (200°C, 5V) | | 68 |

TABLE IV. High temperature UV photodetectors from other wide bandgap materials.

| Material | Photodetector Architecture | Electrode Material | Temperature (°C) | PDCR | Responsivity (A/W) | Notes | Reference |
|---|---|---|---|---|---|---|---|
| ZnO | MSM | Al | 200 | | 9 (RT) 2 (200°C) | Responsivity decreases with increasing temperature due to bandgap shrinkage and lattice scattering. | 72 |
| B-Ga$_2$O$_3$ | MSM | IZO | 427 | 14 (RT, 10V) 1.5 (427°C, 10V) | 0.00032 (RT, 10V) | | 71 |
| Ga$_2$O$_3$ Nanobelt | Photoconductor | Au | 160 | 72,463 (55°C, 20V) 8,892 (160°C, 20V) | 870 (55°C, 20V) 650 (160°C, 20V) | | 75 |
| Diamond | MSM | Ag | 300 | | | UV/visible (200-800nm) discrimination of four orders of magnitude up to 300°C. | 81 |
| BNNS | | Au | 400 | | 9 µA/W (RT, 0V) | Photocurrent increased by a factor of four from RT to 400°C. Thermal noise increased by a factor of 3 from RT to 400°C. | 82 |
| SiCBN | MSM | Al | 200 | 5.5 (RT) 2.5 (200°C) | | | 9 |
| SiCN | MSM | Au | 200 | 6.5 (RT) 2.3 (200°C) | | | 84 |
| n-SiCN/i-SiCN/p-SiCN | n-i-p | Au | 200 | 3180 (RT, -5V) 135.65 (200°C, -5V) | 0.14 (RT, -5V) | QE: 67% (RT, -5V) | 85 |
| n-SiCN/i-SiCN/p-Si | n-i-p | Ni | 200 | 60 (RT, -5V) 4.7 (200°C, -5V) | | | 85 |

| n-SiCN/p-SiCN | n-p | Ni<br>Al | 175 | 1940 (RT, -5V)<br>96.3 (175°C, -5V) | 86 |